\documentclass{article}
\usepackage{amsmath,amsfonts,amssymb}
\usepackage{graphicx}
\usepackage{setspace}
\usepackage{tocloft}
\usepackage{hyperref}
\usepackage{array}
\usepackage{algorithm}
\usepackage{algorithmic}
\usepackage{tikz}
\usepackage{pgfplots}
\usepackage{pgfplotstable,booktabs}
\usepackage{authblk}
\usetikzlibrary{calc,math}
\pgfplotsset{compat=newest}
\usetikzlibrary{arrows,chains,matrix,positioning,scopes}

\newenvironment{customlegend}[1][]{
	\begingroup
	\csname pgfplots@init@cleared@structures\endcsname
	\pgfplotsset{#1}
}{
	\csname pgfplots@createlegend\endcsname
	\endgroup
}
\def\addlegendimage{\csname pgfplots@addlegendimage\endcsname}
\pgfkeys{/pgfplots/number in legend/.style={
		/pgfplots/legend image code/.code={
			\node at (0.295,-0.0225){#1};
		},
	},
}

\newenvironment{keywords}
{\begin{trivlist}\item[]{\bfseries Keywords:}\ }
	{\end{trivlist}}

\newenvironment{AMS}
{\begin{trivlist}\item[]{\bfseries AMS:}\ }
	{\end{trivlist}}

\cftpagenumbersoff{figure}
\cftpagenumbersoff{table}

\graphicspath{{.}{img}}
\newcommand{\review}[1]{{\color{black}#1}}

\begin{document} 
\title{Performance of an iterative wavelet reconstructor\\for the Multi-conjugate Adaptive Optics RelaY\\of \review{ESO's ELT}}

\author[1]{Bernadett Stadler}
\author[1]{Ronny Ramlau}
\affil[1]{Johannes Kepler University, Industrial Mathematics Institute, Altenbergerstraße 69, Linz, Austria, 4040}

\maketitle

\begin{abstract}
The Multi-conjugate Adaptive Optics RelaY (MAORY) is one of the key Adaptive Optics (AO) systems on the European Southern Observatory's Extremely Large Telescope. MAORY aims to achieve a good wavefront correction over a large field of view, which involves a tomographic estimation of the 3D atmospheric wavefront disturbance. Mathematically, the reconstruction of turbulent layers in the atmosphere is severely ill-posed, hence, limits the achievable reconstruction accuracy. Moreover, the reconstruction has to be performed in real-time at a few hundred to one thousand Hertz frame rates. Huge amounts of data have to be processed and thousands of actuators of the deformable mirrors have to be controlled by elaborated algorithms. Even with extensive parallelization and pipelining, direct solvers, such as the Matrix Vector Multiplication (MVM) method, are extremely demanding. Thus, research in the last years shifted into the direction of iterative methods. In this paper we focus on the iterative Finite Element Wavelet Hybrid Algorithm (FEWHA). The key feature of FEWHA is a matrix-free representation of all operators involved, which makes the algorithm fast and enables on the fly system updates whenever parameters at the telescope or in the atmosphere change. We provide a performance analysis of the method regarding quality and run-time for the MAORY instrument using the AO software package COMPASS.
\end{abstract}

\begin{keywords}
{atmospheric tomography, iterative solvers, Extremely Large Telescopes, MAORY, real-time computing}
\end{keywords}

\begin{AMS}
	{65R32, 85-08, 85-10}
\end{AMS}

\section{Introduction}
The Extremely Large Telescope (ELT), which is currently built by the European Southern Observatory (ESO), will become the largest optical/near-infrared telescope in the world and will feature several instruments. One of these instruments is the Multi-conjugate Adaptive Optics RelaY (MAORY)\cite{MAORY}, which acts as an adaptive optics (AO) module. MAORY will not make observations by itself, rather it will enable other instruments, such as the imaging camera MICADO, to take images with an exceptional quality. In this paper, we focus on the MAORY Multi Conjugate Adaptive Optics (MCAO) mode, i.e., the data obtained from several wavefront sensors (WFSs) are utilized to tomographically estimate the 3D atmospheric wavefront disturbance. The usage of multiple guide stars and deformable mirrors (DMs) together with the 3D atmospheric reconstruction enables to correct for multiple directions and a wider field of view. Mathematically, the atmospheric tomography problem is severely ill-posed, i.e., there is an unstable relation between measurements and the solution\cite{Davison83,Nat86}. As a consequence, regularization techniques are required. A common way to regularize this problem is the Bayesian framework, because it allows to incorporate statistical information about turbulence and noise. The random variables are typically assumed to be Gaussian, therefore, the maximum a posterior (MAP) estimate is an optimal point estimate for the solution\cite{Hammer02,Andersen06,RiElFl00,Puech08,DBB10}.

Developing an AO control system for \review{ESO's ELT} is an ambitious and critical task, since an unprecedented amount of data has to be processed in real-time. In order to achieve good results, the implementation of an efficient reconstructor on a high performance computing architecture is inevitable. So far, the standard solver for the atmospheric tomography problem is the Matrix Vector Multiplication (MVM) method. This algorithm precomputes the (regularized) generalized inverse of the system operator in soft real-time and applies a matrix-vector multiplication with the vector of sensor measurements in hard real-time. Even with extensive parallelization and pipelining such a direct solution method is very demanding. Thus, in recent years several iterative solvers have been developed, which are fast and benefit from on the fly system updates. There exist many solvers that are dealing with the atmospheric tomography problem, either directly or iteratively\cite{Fusco,ElGiVo02,GiElVo02b,GiElVo03,YaVoEl06,GiElVo07,GiEl08,RoCoGrScFu10,ThiTa10,Tallon_et_al_10,RaRo12,RoRa13,RaObRoSa13,SaRa15,RafRaYu16}. In this paper we focus on a wavelet based iterative method called the Finite Element Wavelet Hybrid Algorithm (FEWHA)\cite{Yu14,YuHeRa13,YuHeRa13b}. The key feature of FEWHA is the dual domain discretization approach, which leads to a sparse representation of the system matrices and allows an efficient matrix-free representation. This results in a significant reduction of floating point operations (FLOPs) and memory resources. The MAP estimate is computed using a preconditioned conjugate gradient (PCG) method. In order to be able to run FEWHA in real-time, we apply several techniques that reduce the number of PCG iterations. A very common approach in this regard is the warm restart technique, which reuses the solution from the previous time step as initial guess for the PCG method of the next time step. Moreover, we apply preconditioning\cite{YuHeRa13} and an augmented Krylov subspace method\cite{RaSt2021}. \review{A mathematical analysis regarding Krylov subspace methods for FEWHA and first simulations results have been covered in our previous paper\cite{RaSt2021}. The simulations there are performed using the in-house software package MOST\cite{Au17}. This simulation tool has been developed as an alternative to more sophisticated simulators like OCTOPUS\cite{OCTOPUS}, COMPASS\cite{COMPASS} or PASSATA\cite{PASSATA}. It allows to quickly test new approaches, but plenty of simplifications are used and the results are less trustful. In this paper we focus on the performance analysis of FEWHA for the MAORY instrument using COMPASS. We provide a detailed study of the quality as well as the run-time on real-time hardware. Real-time implementations for \review{ESO's ELT} AO system have been previously studied \cite{GPURTC, FPGARTC, MAORY, AOrealtime, GPURTC2, GPURTC3}. Suitable architectures have been evaluated within the Greenflash project\cite{Greenflash1, Greenflash2}. Based on these investigations together with our previously performed analysis\cite{stadler2020realtime} we focus here on a parallel implementation of the algorithm on a multi-core Central Processing Unit (CPU).} 

The paper is organized as follows: We start with a short overview of the atmospheric tomography problem and the Bayesian framework for regularization in Section~\ref{sec:atmo_tomo}. Afterwards, we recall FEWHA including the augmentation and preconditioning concept in Section~\ref{sec:solver}. The quality and computational performance of the algorithm is demonstrated by numerical simulations in Section~\ref{sec:numerical}. Finally, in Section~\ref{sec:conclusion} we state our conclusion.

\section{Atmospheric tomography}\label{sec:atmo_tomo}
In atmospheric tomography we consider a layered model of the atmosphere, where we assume that all turbulences are located at a finite number $L$ of infinitely thin layers $\phi = (\phi_1, ..., \phi_L)$. We aim to reconstruct these turbulent layers, i.e., the refractive index of the turbulent atmosphere, using measurements obtained from WFSs\cite{RoWe96}. The atmospheric tomography operator $A$ relates WFS measurements and layers by
\begin{equation}\label{eq:atmo}
s = (s_g^x, s_g^y)_{g=1}^G = A\phi,
\end{equation}
where $G$ is the number of guide stars, $\phi = (\phi_1, ..., \phi_L)$ denote the $L$ turbulent layers of the atmosphere and $s$ the WFS measurements.

We assume the usage of a Shack-Hartmann (SH) WFS. Then the tomography operator $A$ is decomposed into a geometric propagation operator $P$ into the direction of the guide star and a SH operator $\Gamma$. For a specific guide star $g$ we obtain

\begin{equation*}
s_g = \Gamma_g P_g \phi \quad \text{ for } g=1,...,G.
\end{equation*}

For a SH WFS the vertical and horizontal shifts of the focal points determine the average slope of the wavefront over the area of the lens, known as subaperture\cite{ElVo09,PlSh01,Shack71}. Within a subaperture $\Omega_{ij}$ with $i,j=1,\dots,N$ and $N$ denoting the number of subapertures the SH measurements are modelled as the average slopes of the wavefront aberration $\varphi$. We assume that the incoming wavefront aberration $\varphi$ is approximated by a continuous piecewise bilinear function $\varphi_{ij}$. Hence, we obtain the SH measurements in a subaperture $\Omega_{ij}$ by

\begin{equation*}
s_{ij}^x=\frac{(\varphi_{i,j+1}-\varphi_{i,j})+(\varphi_{i+1, j+1} - \varphi_{i+1,j})}{2},
\end{equation*}

\begin{equation*}
s_{ij}^y=\frac{(\varphi_{i+1,j}-\varphi_{i,j})+(\varphi_{i+1, j+1} - \varphi_{i,j})}{2}.
\end{equation*}

The vectors $s^x$ and $s^y$ are defined as a concatenation of values $s_{ij}^x$ and $s_{ij}^y$ for the set of indices $(i,j)$ that belongs to an active subaperture $\Omega_{ij}$. The SH operator $\Gamma$ maps wavefronts $\varphi$ to SH-WFS measurements $s$ and is given by

\begin{equation*}
s = \begin{pmatrix}s^x\\s^y\end{pmatrix}=\begin{pmatrix}\Gamma^x\varphi\\\Gamma^y\varphi\end{pmatrix} = \Gamma \varphi.
\end{equation*} 

Assuming a layered atmospheric model, the wavefront aberrations in the direction $\theta$ of a natural guide star (NGS) are given by

\begin{equation*}
\varphi_\theta(x)=(P_\theta^{NGS}\phi)(x):=\sum_{\ell=1}^L\phi_l(x+\theta h_\ell),
\end{equation*}

where $\phi_\ell$ is the turbulent layer at altitude $h_\ell$ for $\ell=1,...,L$. We call $P_\theta^{NGS}$ the geometric propagation operator in the direction of the NGS.\\ A laser guide star (LGS) is considered to be at a finite height $H$. Note that we consider here so called sodium LGSs. The incoming wavefront aberrations in the direction $\theta$ of an LGS are given by

\begin{equation*}
\varphi_\theta(x)=(P_\theta^{LGS}\phi)(x):=\sum_{\ell=1}^L\phi_l((1-\frac{h_\ell}{H})x+\theta h_\ell),
\end{equation*}

where $P_\theta^{LGS}$ is called the geometric propagation operator in the direction of the LGS.

The atmospheric tomography problem is a limited angle problem. Mathematically, Equation~\eqref{eq:atmo} is ill-posed, i.e., the relation between the solution and the measurements is unstable\cite{Davison83,Nat86,RaNe17}. To handle this inverse problem regularization is required. Because the Bayesian framework allows the incorporation of statistical information about turbulence and noise, it is frequently used in the community of AO. In this statistical approach we assume $S$ and $\Phi$ to be random variables corresponding to the SH WFS measurements and turbulence layers, respectively. Moreover, we assume additive noise, modeled by the random variable $\eta$. The random variables $\Phi$ and $\eta$ are modeled by Gaussian variables with zero mean and covariance matrices $C_{\Phi}$ and $C_{\eta}$, respectively \cite{Roddier}. The layers are statistically independent, hence, the covariance matrix $C_{\Phi}$ has a block diagonal structure 

\begin{equation*}
C_{\Phi} = diag(C_1, \dots, C_L).
\end{equation*}

\review{We assume that the noise is identically distributed in each subaperture and that the $x$- and $y$-measurements are uncorrelated. The covariance matrix for an NGS direction $g$ is then defined by
	
\begin{equation}\label{eq:NGS-noise}
C_g = \sigma^2_{\text{NGS}} I,
\end{equation}

where $\sigma^2_{\text{NGS}}=n_{photons}^{-1}$ is the noise variance of a single measurement and $n_{photons}$ denotes the number of photons per subaperture.

For LGSs the sodium layer thickness must be considered when modeling the photon noise. The vertical density profile of the laser beam scatter is modeled by a Gaussian random variable with mean $H$ and full width at half maximum (FWHM) of the sodium density profile in meters, given by

\begin{equation*}
FWHM = 2\sqrt{2 ln(2)}\ \sigma_{\text{LGS}}.
\end{equation*} 

The elongation vector in a subaperture $\Omega_{ij}$ is defined as

\begin{equation*}
\beta_{ij} = (\beta_{ij,1}, \beta_{ij,2}) = \frac{FWHM}{H^2}\left((\bar{x}_i,\bar{x}_j)-(x_1^{LL}, x_2^{LL})\right),
\end{equation*}

where $(x_1^{LL}, x_2^{LL})$ are the laser launch positions and $(\bar{x}_i, \bar{x}_j)$ the midpoints of the subaperture $\Omega_{ij}$. The spot elongated noise covariance matrix in a subaperture $\Omega_{ij}$ is given by

\begin{equation}\label{eq:LGS-noise}
C_{ij} = \sigma_{\text{LGS}}^2\left(I + \frac{\alpha_\eta^2}{f^2}\begin{pmatrix}\beta_{ij,1}^2 & \beta_{ij,1}\beta_{ij,2}\\\beta_{ij,1}\beta_{ij,2} & \beta_{ij,2}^2\end{pmatrix}\right),
\end{equation}

where $I$ denotes the identity matrix, $\sigma^2_{\text{LGS}}=n_{photons}^{-1}$ and $f$ is the FWHM of the non-elongated spot. To cope with noise sources that are not included into the model above, e.g., read out noise, we introduce the fine-tuning parameter $\alpha_\eta$. Altogether we obtain the noise covariance matrix

\begin{equation*}
C_{\eta} = diag(C_1, \dots, C_{G_{LGS}},C_{G_{LGS+1}}, \dots, C_{G}),
\end{equation*}

where $G = G_{LGS}+G_{NGS}$ denotes the number of guide stars.} 

For the above described setting the maximum a posteriori (MAP) estimate provides an optimal point estimate for the solution\cite{Fusco}, which is given by the solution of the linear system of equations

\begin{equation}\label{eq:MAP_atmo}
(A^* C_\eta^{-1}A + C_\phi^{-1})\phi = A^* C_\eta^{-1}s.
\end{equation}

Here $A^*$ denotes the adjoint tomography operator. Note, that the dimension of the operator $A$ is, in general, larger for bigger telescopes. Hence, solving the atmospheric tomography problem for \review{ESO's ELT} in real-time is a highly non-trivial task.

\section{A wavelet based iterative solver for MAORY}\label{sec:solver}
The trade-off between optimal performance and computational complexity for the atmospheric tomography problem of ELTs has triggered the development of iterative real-time reconstructors with a complexity of $\mathcal{O}(n)$ operations. Most of them still rely on the formulation of the forward problem as a matrix equation, i.e., the matrix has to be assembled frequently during the observation of one scientific object. To overcome this limitation, the Finite Element Wavelet Hybrid Algorithm (FEWHA) has been proposed\cite{Yu14,YuHeRa13,YuHeRa13b,RaSt2021}. FEWHA utilizes a matrix-free representation of all operators involved and a conjugate gradient (CG) based approach to compute the solution of Equation~(\ref{eq:MAP_atmo}). Moreover, several techniques are applied to reduce the number of PCG iterations, and thus the run-time, to a minimum\cite{YuHeRa13,RaSt2021}.

\subsection{Dual domain discretization}
In order to numerically compute a solution of Equation~\eqref{eq:MAP_atmo}, discretization is required. There are certain fundamental advantages of using wavelets, as already extensively studied for FEWHA\cite{YuHeRa13b}. The main idea is to use compactly supported orthonormal wavelets for representing the turbulent layers. Wavelets allow in particular a diagonal approximation of $C_\phi$. However, the atmospheric tomography operator $A$ has a more efficient representation in a finite element domain, where continuous piecewise bilinear functions are utilized to represent wavefronts and layers. As the discrete wavelet transform is of complexity $\mathcal{O}(n)$, this allows to evaluate $C_\phi$ in the wavelet domain and $A$ in a the bilinear domain.

We utilize a square grid with equidistant spacing on the subaperture domain at the telescope pupil $\Omega$ to define the piecewise bilinear wavefront functions. The piecewise bilinear layer functions are defined using a square mesh with equidistant spacing on $\Omega_\ell$, the domain on which the turbulent layer $\ell$ is defined. This mesh consists of $2^{2J_\ell}$ points, where $J_\ell$ denotes the number of wavelet scales. Utilizing this dual domain discretization approach \cite{YuHeRa13b} we obtain the discretized version of the MAP equation as

\begin{equation}\label{eq:fewha}
(\boldsymbol{W}^{-T}\hat{A}^TC_\eta^{-1}\hat{A}\boldsymbol{W}^{-1} + \alpha D)c=\boldsymbol{W}^{-T}\hat{A}^TC_\eta^{-1}s,
\end{equation}

where $\hat{A}$ is the atmospheric tomography operator in the finite element domain and $\hat{A}^T$ is the transposed matrix. We denote the linear mapping between the finite element and the wavelet domain by $\boldsymbol{W} = diag(\delta_1W,...,\delta_LW)$, where $W$ is the discrete wavelet transform and $\delta_l$ is the scaling constant at layer $\ell$. The operator $C_\eta^{-1}$ denotes the inverse covariance matrix of the noise and $D$ is a diagonal approximation of $C_\phi^{-1}$ in the frequency domain. We introduce a scalar factor $\alpha$ for tuning the balance between the fitting and the regularizing terms. If $\alpha$ is increased, the modeling error is stabilized, but the quality of the reconstruction is reduced. In practice, $\alpha$ is tuned by hand via numerical simulations for a specific test configuration. The vector $c$ is a concatenation of all wavelet coefficients of all turbulence layers and the vector $s$ is the concatenation of all SH WFS measurements from all guide star directions. \review{In all our simulations we use periodic Daubechies-$N$ wavelets with $N=3$, which are an orthogonal wavelet family with compact support. A compactly supported orthonormal wavelet basis has favorable properties in the frequency and the spatial domains. The properties in the frequency domain enable to approximate $C_\phi$ by a diagonal matrix. Note that the larger $N$ is the more computations have to be performed to apply the discrete wavelet transform. Based on our numerical tests we believe that $N=3$ is the optimal choice for balancing between smoothness and run-time. }

For the sake of simplicity, we define the left-hand side operator of Equation~\eqref{eq:fewha} by

\begin{equation}\label{eq:M}
M := (\boldsymbol{W}^{-T}\hat{A}^T C_\eta^{-1}\hat{A}\boldsymbol{W}^{-1} + \alpha D)
\end{equation}

and the right-hand side as

\begin{equation*}\label{eq:b}
b := \boldsymbol{W}^{-T}\hat{A}^T C_\eta^{-1}s.
\end{equation*}

Note that the matrix $M \in \mathbb{R}^{2^{2J_{\ell}}L\times 2^{2J_{\ell}}L}$ is symmetric and positive definite. Hence, we can solve Equation~\eqref{eq:fewha} using the iterative preconditioned CG (PCG) method. 

\review{\subsection{Tip-tilt uncertainty}
	LGSs introduce a tip-tilt uncertainty, which has to be taken into account in order to achieve a good correction. For FEWHA the incorrect tip-tilt component is removed directly in Equation~\eqref{eq:fewha}. Let us denote by $\mathcal{M}$ the SH WFS mask associated to an LGS. We define the two tip-tilt measurement vectors of dimension $2|\mathcal{M}|$ by
	
	\begin{equation*}
	t^x=(e \quad \mathbf{0})^T \quad\quad t^y=(\mathbf{0} \quad e)^T,
	\end{equation*}
	
	where $e=(1,\dots,1)^T$ denotes a vector of ones and $\mathbf{0}$ a vector of zeros, both of dimension $|\mathcal{M}|$. The tip-tilt projection operator $T$ is then given by
	
	\begin{equation*}
	T=\frac{1}{|\mathcal{M}|}(t^x \quad t^y)(t^x \quad t^y)^T.
	\end{equation*}
	
	In order to remove the incorrect tip-tilt we apply operator $T$ to the inverse noise covariance matrix $C_\eta$ as defined in Equation~\eqref{eq:LGS-noise}. In fact, $C_\eta$ is modified for each LGS direction $g=1,\dots, G_{LGS}$ to
	
	\begin{equation*}
	\hat{C}_g^{-1}=(I-T) C_g^{-1}(I-T),
	\end{equation*}
	
	where $I$ is the identity matrix and the operator $T$ applies an orthonormal projection into the measurement space of tip and tilt.}

\subsection{Convergence improvements}
Within the control of an AO system we are dealing with several right-hand sides, corresponding to different WFS measurements, available consecutively in every time step. We introduce the following notation for Equation~\eqref{eq:fewha} for several time steps $i=1,2,\dots$ 

\begin{equation*}\label{eq:severalRHSAO}
M c^{(i)}=b^{(i)}.
\end{equation*}

Iterative solvers, such as FEWHA or the Fractal Iterative Method\cite{Tallon_et_al_10}, have to be reapplied for every single time step $i$. This is costly in terms of computational speed. Especially compared to a direct solver where the factorization, which is computed in soft real-time, can be reused independently of the right-hand side.

A very common technique to reduce the number of PCG iterations is preconditioning. We utilize a modified Jacobi preconditioner in which the low and high frequencies are weighted differently\cite{YuHeRa13}. The classical Jacobi preconditioner is a diagonal matrix given by $J=diag(M)$, hence, very easy to invert and efficient to apply. The benefit of a Jacobi preconditioner is the reduction of CG iterations and an increased stability and robustness of the whole method. However, a standard Jacobi preconditioner dampens the high scales too much in comparison with the lower ones. The high and low wavelet scales are related to high and low frequency regimes of the atmospheric layers\cite{YuHeRa13}. We introduce the parameter $\tau$ in order to balance the level of damping and use a slightly modified form

\begin{equation}\label{eq:precond}
J =diag((\boldsymbol{W}^{-T}\hat{A}^T\hat{C}_\eta^{-1}\hat{A}\boldsymbol{W}^{-1}) + \alpha \max(D,\tau I)).
\end{equation} 

Here $I$ denotes the identity matrix and $\tau$ is a non-negative scalar factor. Note that the maximum value of the two matrices is taken component wise. If we choose $\tau=0$ we arrive at the standard Jacobi preconditioner, whereas for a very large $\tau$ the term $\tau I$ dominates.

The sensor measurements obtained by the SH WFSs do not change significantly from time step to time step, which also holds for the right-hand side $b^{(i)}$. A common procedure within AO which exploits this behavior is to utilize the solution from the previous time step $c^{(i)}$ as initial guess for the PCG method of the next time step. This is often referred to as warm restart. Going one step further into that direction, we additionally reuse the search directions from the previous time step. We store these search directions in a matrix $P_m^{(i)}$, where $m$ denotes the number of PCG iterations. In fact, we extend FEWHA with a so called augmented Krylov subspace approach\cite{RaSt2021}. The augmented PCG algorithm requires some additional computations compared to the classical algorithm, such as scalar products and vector updates. However, compared to the overall number of operations these calculations are negligible. Moreover, the number of PCG iterations $m$ for our numerical simulations is very small (below $5$). Hence, the additional memory requirements for storing the search directions for one time step $i$ in $P_m^{(i)}$ are minor.

\subsection{The algorithm}
The general structure of the wavelet reconstructor for one time step $(i+1)$ is outlined in Algorithm~\ref{alg:augFEWHA}\cite{RaSt2021}. The input parameters of the algorithm are: the measurement vector $s^{(i+1)}$, corresponding either to open or closed loop measurements, the solution from the previous time step $c^{(i)}$, which acts as initial guess for the augmented PCG algorithm, and the previous right-hand side and residual $b^{(i)}$ and $r^{(i)}$. Moreover, we use the current actuator commands $a^{(i)}$ in combination with the previous ones $a^{(i-1)}$ for applying closed loop control. The maximum number of PCG iterations is fixed to $\mathrm{iter}$ iterations. This value is determined via numerical simulations; see Section~\ref{sec:numerical}. The augmented PCG method requires the descent directions of the previous time step $P^{(i)}$ as input. To avoid unnecessary recomputations we further save $M$ applied to these search directions and denote this matrix by $Q^{(i)}$. The output is the new vector of actuator commands $a^{(i+1)}$, used by the control scheme to deform the adaptive mirror.

An AO system can operate either in closed or in open loop. If we apply open loop control, the measurements are directly obtained from the wavefronts. If we use a closed loop control, the pseudo open loop measurements have to be calculated as a first step of the algorithm; see Line~\ref{line:openLoop}. Due to a two-step delay\cite{Poettinger_2019} we use the actuator commands $a^{(i-1)}$ from time step $(i-1)$ to compute the pseudo open loop measurements $s^{(i+1)}$. The right-hand side $b^{(i+1)}$ is computed in Line~\ref{line:RHS} with the new measurement vector $s^{(i+1)}$, and subsequently the initial residual $r_0^{(i+1)}$ is updated in Line~\ref{line:res}. The atmospheric reconstruction takes place in Line~\ref{line:PCG}, where $P^{(i)}$ and $Q^{(i)}$ are used within the augmented PCG method to decrease the number of iterations by projection. In Line~\ref{line:fitting} the layers are first transformed back from the wavelet into the finite element domain by applying the inverse discrete wavelet transform. Subsequently, the mirror shapes $\tilde{a}$ are fitted to the reconstructed atmosphere by applying the mirror fitting operator $F$. This operator is different for each AO system\cite{RoRa12}. Closed or open loop control is applied in Lines~\ref{line:closedcontrol}-\ref{line:opencontrol}. The new actuator commands are calculated as a linear combination of the current and the reconstructed actuator commands, weighted by a scalar value $\mathrm{gain} \in [0,1]$. This gain control improves the stability of the reconstruction. For closed loop control the artificially added DM shapes $a^{(i-1)}$ are subtracted from the computed mirror shapes $\tilde{a}$. 

\begin{algorithm}
	\caption{Augmented wavelet reconstructor \cite{RaSt2021}}
	\small
	\label{alg:augFEWHA}
	\begin{algorithmic}[1]
		\STATE{\textbf{Input:}\quad\quad~$s^{(i+1)}=(s_g)^G_{g=1}$ (measurement vector)\\
			\quad\quad\quad\quad\quad~$\mathrm{gain}$ (scalar weight)\\
			\quad\quad\quad\quad\quad~$c^{(i)}$ (previous wavelet coefficients)\\
			\quad\quad\quad\quad\quad~$b^{(i)}$ (previous right-hand side)\\
			\quad\quad\quad\quad\quad~$r^{(i)}$ (previous residual vector)\\
			\quad\quad\quad\quad\quad~$a^{(i-1)}, a^{(i)}$ (previous two DM shape)\\
			\quad\quad\quad\quad\quad~$\mathrm{iter}$ (maximum number of PCG iterations)\\
			\quad\quad\quad\quad\quad~$J^{-1/2}$ (Jacobi preconditioner)\\
			\quad\quad\quad\quad\quad~$P^{(i)}$,$Q^{(i)}$ (previous descent directions)
		}
		\STATE{\textbf{Output:}\quad~$a^{(i+1)}$ (actuator commands)}
		
		\IF{loop = closed} \label{line:openLoop}
		\STATE $s^{(i+1)} = s^{(i+1)} + \Gamma a^{(i-1)}$
		\ENDIF
		\vspace{0.3cm}
		
		\STATE $b^{(i+1)} = \mathbf W^{-T} \hat{A}^T \hat{C}_{\eta}^{-1} s^{(i+1)}$ \label{line:RHS}
		\STATE $r_0 = b^{(i+1)} - M c^{(i)} = (b^{(i+1)} - b^{(i)}) + r^{(i)}$ \label{line:res} 	
		\vspace{0.3cm}				
		\STATE $[c^{(i+1)}, r^{(i+1)}, P^{(i+1)}, Q^{(i+1)}] = augPCG(c^{(i)}, r_0, J^{-1/2}, P^{(i)}, Q^{(i)}, \mathrm{iter})$\\\label{line:PCG}
		\vspace{0.3cm}
		\STATE$\tilde a = F \mathbf W^{-1} c^{(i+1)}$\label{line:fitting}
		
		\vspace{0.3cm}
		\IF{loop = closed}\label{line:closedcontrol}
		\STATE $a^{(i+1)} = a^{(i)} + \mathrm{gain} \cdot (\tilde a - a^{(i-1)})$ 
		\ELSIF{loop = open}
		\STATE $a^{(i+1)} = (1 - \mathrm{gain}) \cdot a^{(i)} + \mathrm{gain} \cdot \tilde a$ 
		\ENDIF \label{line:opencontrol}
	\end{algorithmic}
\end{algorithm}

For more details on the augmented Krylov subspace method within FEWHA we refer to our paper about an augmented wavelet reconstructor for atmospheric tomography\cite{RaSt2021}.

\section{Numerical simulations}\label{sec:numerical}
For our numerical simulations we use the software package COMPASS\cite{COMPASS}, which allows in particular to simulate all critical subcomponents of an AO system in the context of \review{ESO's ELT}. The tool takes advantage of the GPU hardware architecture, and thus is able to provide an adequate execution speed for large simulations. FEWHA is not included into COMPASS. The communication between the algorithm and the simulation tool is handled by data exchange via the file system. As a benchmark for the quality evaluation we use the COMPASS internal least-squares (LS) reconstructor, for which the control matrix is computed via a singular value decomposition and pseudo-inversion of the interaction matrix. \review{Note that we are not able to compare our results with the COMPASS internal MVM method, because the NVIDIA Tesla V100 GPU we are using has too little memory resources to run this matrix-based approach. For a quality evaluation of the MVM algorithm using the PASSATA simulator we refer to\cite{MAORY_AO_Performance}.}

\subsection{System configuration}
We simulate a telescope that gathers light through a primary mirror of $37$~m diameter, where approximately $11~\%$ of the mirror are obstructed. We utilize a $35$~layer atmosphere\cite{layer_profile} that follows the von Karman statistics. In order to avoid the mirror fitting step we reconstruct $3$ layers directly at the altitude of the DMs. Then the operator $F$ in Line~\ref{line:fitting} of Algorithm~\ref{alg:augFEWHA} becomes the identity matrix. \review{Note that the algorithm is capable of reconstructing more layers, which would improve the reconstruction quality. However, in terms of run-time this is not feasible for the MAORY setting, as it requires to solve an additional minimization problem for the fitting step.} We evaluate the quality using the Strehl ratio in the K band, i.e., at a wavelength of $2200$~nm. The system parameters are summarized in Table~\ref{tab:general_setting}. Details about the reconstructed layers can be found in Table~\ref{tab:layers}. 

\begin{table}
	\centering
	\renewcommand{\arraystretch}{1.3}
	\begin{tabular}{|r|c|}
		\hline
		\textbf{Parameter} & \textbf{Value}\\\hline\hline
		Telescope diameter &  $37$~m\\\hline
		Central obstruction & $11\%$\\\hline
		Na-layer height & $90$~km\\\hline
		Na-layer FWHM & $11.4$~km\\\hline
		Outer scale $L_0$ & $25$~m\\\hline
		Field of View & $1$~arcmin\\\hline
		Simulated duration & $1$~s\\\hline
		Delay & $2$~frames\\\hline
		Evaluation criterion & LE Strehl\\\hline
		Evaluation wavelength & K band ($2200$~nm) \\\hline
	\end{tabular}
	\caption{System parameters.}
	\label{tab:general_setting}
\end{table}

\begin{table}
	\centering
	\renewcommand{\arraystretch}{1.3}
	\begin{tabular}{|c|c|c|c|c|c|}
		\hline
		\textbf{Layer} & \textbf{Altitude} & \textbf{Strength} & \textbf{Scales $J_\ell$} & \textbf{Grid points} & \textbf{Spacing $\delta_\ell$} \\\hline\hline
		$1$ & $0$~m & $0.75$ & $7$ & $128 \times 128$ & $0.5$~m\\\hline
		$2$ & $4000$~m & $0.15$ & $6$ & $64 \times 64$ & $1.0$~m\\\hline
		$3$ & $12700$~m & $0.1$ & $6$ & $64 \times 64$ & $1.0$~m\\\hline
	\end{tabular}
	\caption{Reconstructed layer configuration.}
	\label{tab:layers}
\end{table}

We focus here on an MCAO mode, which uses $3$ DMs as defined in Table~\ref{tab:DMs}. \review{We assume an equidistant actuator spacing for all DMs\cite{Ca12}. The shape of the mirror is determined via bilinear interpolation between the actuators.} Note that within FEWHA the NGS and the LGS problem are coupled, i.e., we do not use an additional tip-tilt mirror. For the COMPASS internal LS reconstructor we use an additional TT mirror in our simulations.

\begin{table}
	\centering
	\renewcommand{\arraystretch}{1.3}
	\begin{tabular}{|r|c|c|c|}
		\hline
		\textbf{Parameter} & \textbf{M4} & \textbf{DM1} & \textbf{DM2}\\\hline\hline
		Number of actuators & $75\times 75$ &  $47\times 47$ & $37\times 37$\\\hline
		DM altitude & $0$~km & $4$~km & $12.7$~km\\\hline
		DM actuator spacing & $0.5$~m & $1$~m & $1$~m\\\hline
	\end{tabular}
	\caption{DM configuration.}
	\label{tab:DMs}
\end{table}

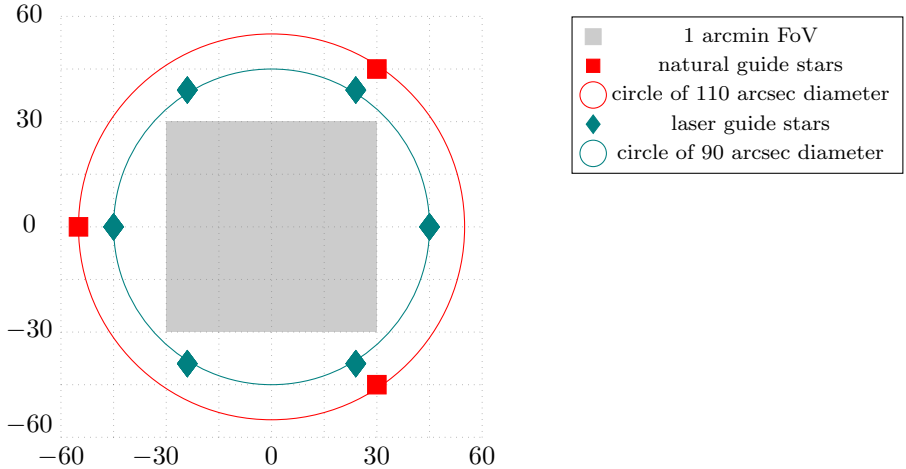
\begin{figure}
	\centering
	\begin{tikzpicture}[scale=1.4]
	\fill[gray!40!white] (2,2) rectangle (4,4);
	\draw [very thin, dotted, gray, step=0.5] (0.9999,0.9999) grid (5,5);
	\draw (0.7,1.3) node[below] {$-60$};
	\draw (0.7,2.2) node[below] {$-30$};
	\draw (0.7,3.2) node[below] {$0$};
	\draw (0.7,4.2) node[below] {$30$};
	\draw (0.7,5.2) node[below] {$60$};
	\draw (1,1) node[below] {$-60$};
	\draw (1.9,1) node[below] {$-30$};
	\draw (3,1) node[below] {$0$};
	\draw (4,1) node[below] {$30$};
	\draw (5,1) node[below] {$60$};
	\draw[teal] (3,3) circle (1.5);
	\draw[red] (3,3) circle (1.8333);
	\foreach \x in {2,2.5,3,3.5,4}
	\draw [teal] plot [only marks, mark size=3.5, mark=diamond*] coordinates {(2.2,4.3) (3.8,4.3) (1.5,3) (4.5,3) (2.2,1.7) (3.8,1.7)};
	\draw [red] plot [only marks, mark size=2.5, mark=square*] coordinates {(1.1667,3) (4,1.5) (4,4.5)};
	\begin{customlegend}[
	legend entries={
		$1$ arcmin FoV,
		natural guide stars,
		circle of $110$ arcsec diameter, 
		laser guide stars,
		circle of $90$ arcsec diameter, 
	},
	legend style={at={(9,5)},font=\footnotesize}]
	\addlegendimage{only marks, mark=square*, color=gray!40!white, mark size=3}
	\addlegendimage{only marks, mark=square*, color=red, mark size=2.5}
	\addlegendimage{only marks, mark=o, color=red, mark size=5}
	\addlegendimage{only marks, mark=diamond*, color=teal, mark size=3.5}
	\addlegendimage{only marks, mark=o, color=teal, mark size=5}
	\end{customlegend}
	\end{tikzpicture}
	\caption{Star asterism of NGSs (red) in a circle of $110$ arcsec diameter and LGSs (teal) in a circle of $1.5$ arcmin diameter. The $1$ arcmin FoV is marked in gray.}
\label{fig:MCAO_asterism}
\end{figure}

The six high order SH WFSs that measure the light coming from the LGSs, are equipped with $74\times 74$ subapertures each consisting of $12\times 12$ pixels. The three low order WFSs which are used for measuring the NGSs aberrations and correcting for the tip-tilt uncertainty are equipped with $2 \times 2$ subapertures, each consisting of $6\times 6$ pixels. The LGSs are positioned in a circle of $90$~arcsec diameter and the NGSs in a circle with a diameter of $110$~arcsec. The MCAO star asterism is shown in Figure~\ref{fig:MCAO_asterism}. We compute the measurements from the slope of the incoming wavefronts via a weighted center of gravity (WCoG) algorithm in COMPASS for the high as well as the low order SH WFSs. \review{During our simulations it turned out that a WCoG with optimized weights provides the best reconstruction quality. Note that we do not use any slope-side gain.} Details about the parameters can be found in Table~\ref{tab:LGS}.

\begin{table}
	\centering
	\renewcommand{\arraystretch}{1.3}
	\begin{tabular}{|r|c|c|}
		\hline
		\textbf{Parameter} & \textbf{LGS-WFS} & \textbf{NGS-WFS}\\\hline\hline
		Type & SH WFS & SH WFS\\\hline
		Number &  $6$ & $3$\\\hline
		Geometry & $74\times 74$ subap. & $2 \times 2$ subap.\\\hline
		Subaperture size & $12\times 12$ pixels & $6\times 6$ pixels \\\hline
		Optical throughput & $0.23$ & $0.33$ \\\hline
		FoV per subaperture & $16.8$~arcsec & $1.3$~arcsec \\\hline
		GS asterism & $90$ arcsec diameter & $110$ arcsec diameter\\\hline
		Wavelength & $589$ nm & $1650$ nm\\\hline
		Detector RON & $3.0~e^-$/pixel/frame & $0.5~e^-$/pixel/frame\\\hline
		Centroiding algorithm & WCoG & WCoG\\\hline
	\end{tabular}
	\caption{WFS configuration.}
	\label{tab:LGS}
\end{table}

\subsection{Quality evaluation}\label{sec:computational}
Several parameters within FEWHA need to be optimized in order obtain a good quality. We perform an optimization of method specific parameters via numerical simulations in COMPASS. The turbulence is simulated according to median seeing conditions with a Fried parameter of $r_0=0.157$\review{~m}. The method specific parameters of FEWHA are the regularization parameter $\alpha$ (see Equation~\eqref{eq:fewha}) and the preconditioner threshold $\tau$ (see Equation~\eqref{eq:precond}).  \review{ We use an integrator control to deal with the time delay between the moment when measurements are acquired by the WFS and the time when the DM correction is applied . To indicate that we are in principle applying the wrong correction we use an output or loop gain (see Line~\ref{line:closedcontrol} of Algorithm~\ref{alg:augFEWHA}). An optimization might be possible, but is not considered in the paper. Note that FEWHA is completely independent from the control strategy.} To cope with noise sources that are not included into the LGS model, e.g., read out noise, we use the fine-tuning parameter $\alpha_\eta$\cite{Bechet}. If $\alpha_\eta=0$, then the model coincides with the NGS model, whereas for $\alpha_\eta=1$ we have the full LGS model. All method parameters are variable and have to be tuned for the specific test setting and noise level. Table~\ref{tab:methodParams} summarizes the optimal parameter values for FEWHA for a highflux test configuration with $10000$ photons per subaperture per frame and a lowflux setting with $500$ photons per subaperture per frame \review{ for LGSs as well as NGSs. Note that the algorithm is capable of handling different number of photons for LGSs and NGSs. The photon flux is taken into account in the noise covariance matrix $C_\eta$; see Equation~\eqref{eq:NGS-noise} and Equation~\eqref{eq:LGS-noise}. } The method reacts particularly sensitive to changes in the regularization parameter. The number of PCG iterations $\mathrm{iter}$ is fixed for a specific test setting and is chosen such that it optimally balances between quality and run-time.

\begin{table}
	\centering
	\renewcommand{\arraystretch}{1.3}
	\begin{tabular}{|rl|c|c|}
		\hline
		\textbf{Description}\hspace{1cm} &  & \textbf{Highflux} & \textbf{Lowflux}\\\hline\hline
		Regularization & $\alpha$ & $80$ & $16$\\
		Spot elongation tuning & $\alpha_\eta$ & $0.2$ & $0.2$\\
		Preconditioner scaling & $\tau$ & $10^5$ & $10^5$\\
		PCG iterations & $\mathrm{iter}$ & $2-4$ & $2-4$\\
		Loop gain & $\mathrm{gain}$ & $0.8$ & $0.6$\\\hline
	\end{tabular}
	\caption{Optimal method parameters for median seeing conditions with a Fried parameter of $r_0=0.157$\review{~m}.}
	\label{tab:methodParams}
\end{table}

\begin{figure}
	\centering
	\includegraphics[width=1.0\textwidth]{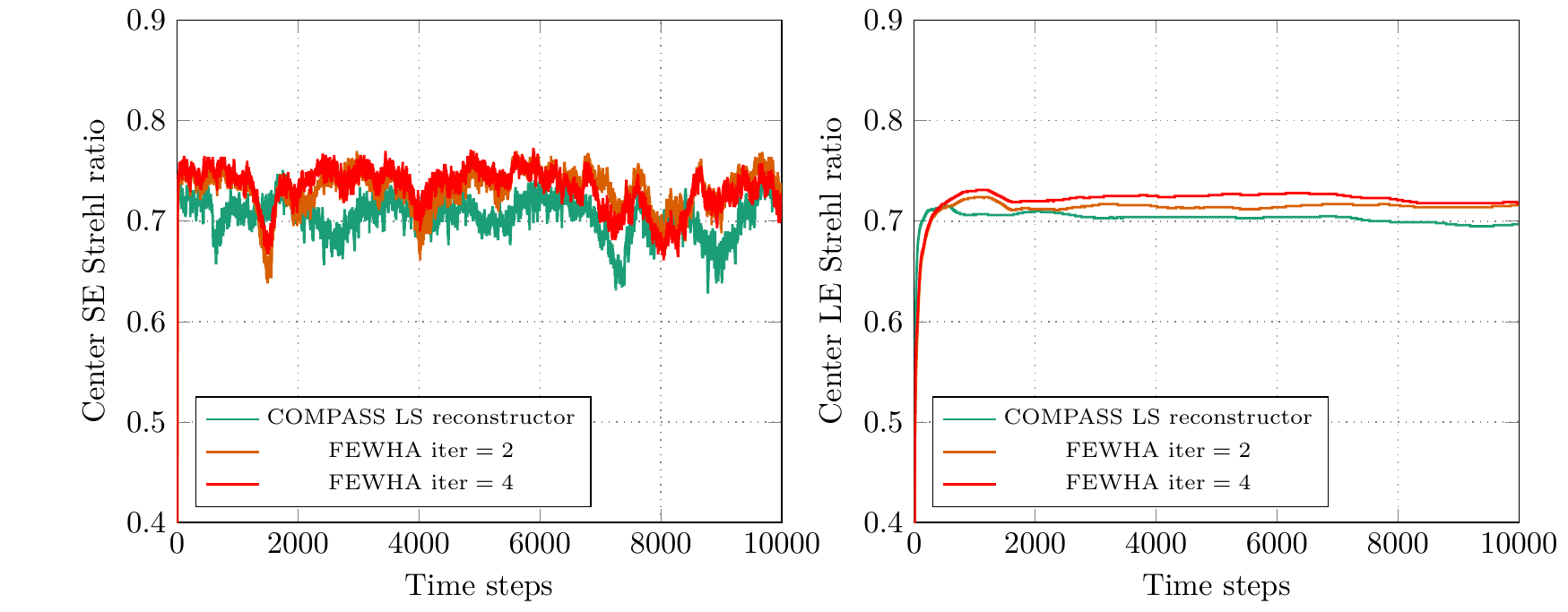}
	\caption{Center SE (left) and LE (right) Strehl ratio for the the COMPASS LS reconstructor (green) and FEWHA with $2$ (orange) and $4$ (red) PCG iterations over $10000$ time steps. Highflux simulation with $10000$ photons per subaperture per frame. The turbulence is simulated according to median seeing conditions with a Fried parameter of $r_0=0.157$\review{~m}.}
	\label{fig:highfluxVsTime}
\end{figure}

We start with an analysis of the highflux setting, using the optimal parameter values from Table~\ref{tab:methodParams}. We compare the quality of the wavelet reconstructor with the COMPASS LS reconstructor in terms of SE and LE Strehl ratio. In Figure~\ref{fig:highfluxVsTime} we show the center SE (left) and LE (right) Strehl ratio over $10000$ time steps. We observe that FEWHA with only $2$ PCG iterations (orange) provides a better quality than the LS reconstructor (green). When using $4$ PCG iterations for FEWHA (red) we can increase the quality even further. Using more than $4$ iterations does not yield additional improvements. Note that the number of PCG iterations is a trade-off between quality and run-time. 

\begin{figure}
	\centering
	\includegraphics[width=0.5\textwidth]{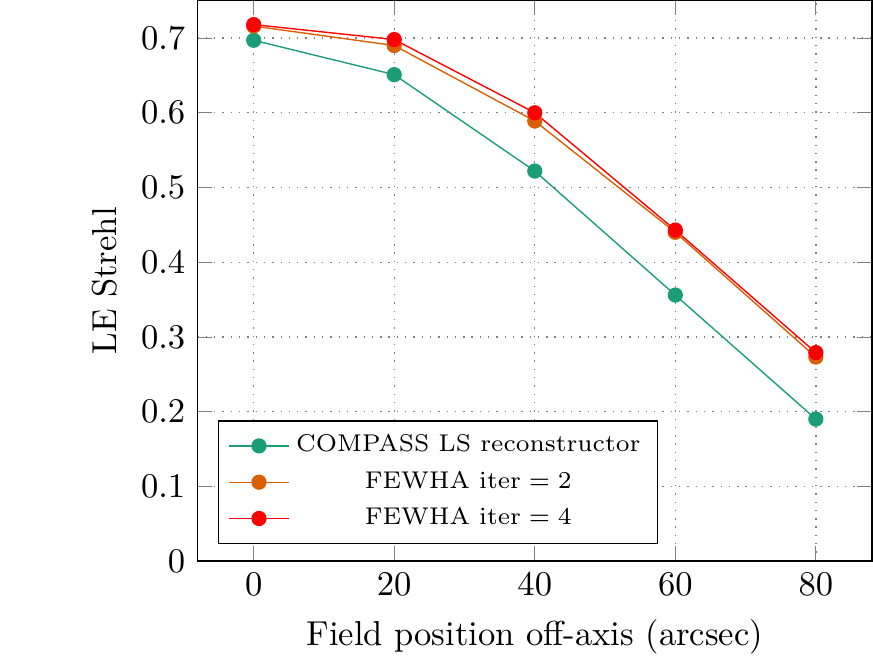}
	\caption{Center LE Strehl ratio after $10000$ time steps of the COMPASS LS reconstructor (green) and FEWHA with $2$ (orange) and $4$ (red) PCG iterations versus the field off-axis position. Highflux simulation with $10000$ photons per subaperture per frame. The turbulence is simulated according to median seeing conditions with a Fried parameter of $r_0=0.157$\review{~m}.}
	\label{fig:highfluxVsSeparation}
\end{figure}

In Figure~\ref{fig:highfluxVsSeparation} we illustrate the LE Strehl ratio versus the field off-axis position for FEWHA with $2$ (orange) and $4$ (red) iterations and the LS reconstructor of COMPASS (green). We observe that FEWHA provides a significant better quality compared to the LS reconstructor for an increasing off-axis position.
  
\begin{figure}
	\centering
	\includegraphics[width=1.0\textwidth]{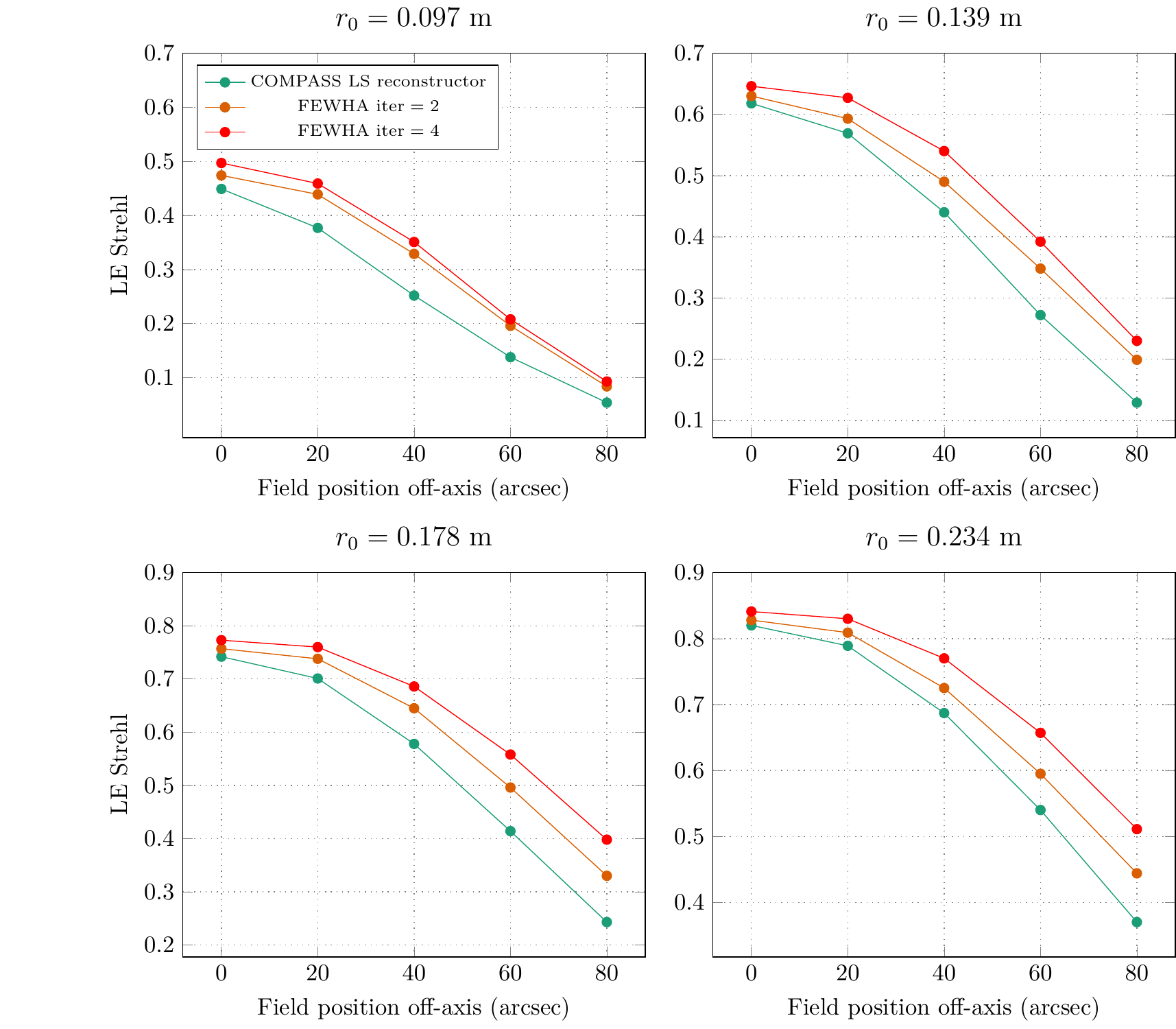}
	\caption{LE Strehl ratio after $10000$ time steps of the COMPASS LS reconstructor (green) and FEWHA with $2$ (orange) and $4$ (red) PCG iterations versus the field off-axis position. Highflux simulation with $10000$ photons per subaperture per frame. The four plots correspond to different Fried parameters $r_0$.}
	\label{fig:highfluxVsSeparationFried}
\end{figure}

To study the performance of the reconstruction methods in more detail we analyze the behavior for different seeing conditions, i.e., for four additional Fried parameters $r_0$. Again we use the highflux configuration and the optimal parameter values from Table~\ref{tab:methodParams}. Note that larger values of $r_0$ correspond to good seeing conditions, whereas smaller values refer to bad seeing and strong perturbations. Hence, a lower Fried parameter leads to a lower LE Strehl ratio. In Figure~\ref{fig:highfluxVsSeparationFried} we illustrate the LE Strehl versus the field off-axis position for four different Fried parameters for the COMPASS LS reconstructor (green) and FEWHA with $2$ (orange) and $4$ (red) PCG iterations. Again we observe that
FEWHA provides a better quality than the LS reconstructor for all off-axis positions. In the center the LE Strehl ratio of FEWHA is slightly better than that of the LS method. For an increasing off-axis angle the difference in quality between FEWHA and the LS reconstructor becomes larger.

\begin{figure}
	\centering
	\includegraphics[width=1.0\textwidth]{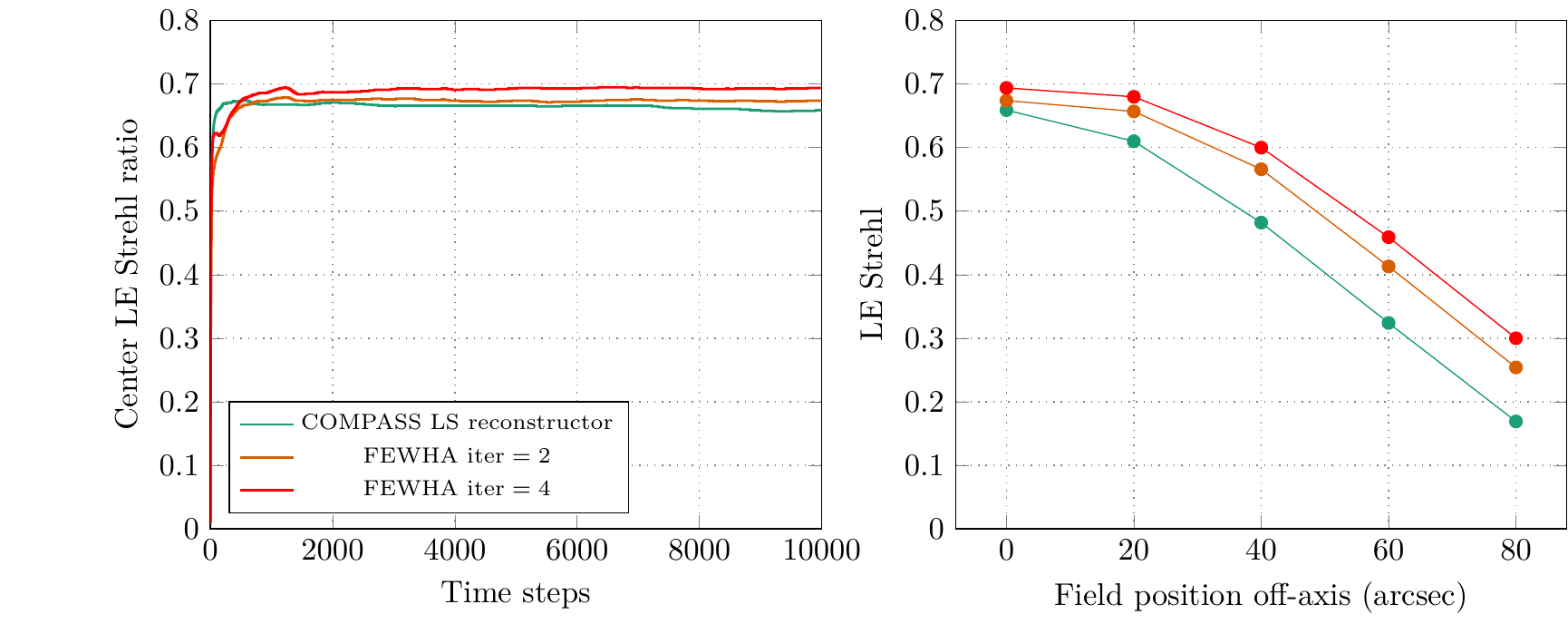}
	\caption{Center LE Strehl ratio over time steps (left) and LE strehl ratio versus the field off-axis position (right) of the COMPASS LS reconstructor (green) and FEWHA with $2$ (orange) and $4$ (red) PCG iterations. Lowflux simulation with $500$ photons per subaperture per frame. The turbulence is simulated according to median seeing conditions with a Fried parameter of $r_0=0.157$\review{~m}.}
	\label{fig:lowflux}
\end{figure}

We continue with an analysis of the lowflux setting with $500$ photons per subaperture per frame. The optimal parameter values for FEWHA are again taken from Table~\ref{tab:methodParams}. In the left plot of Figure~\ref{fig:lowflux} we show the center LE Strehl ratio over $10000$ time steps for FEWHA with $2$ (orange) and $4$ (red) PCG iterations and the LS reconstructor of COMPASS (green). In the right plot of Figure~\ref{fig:lowflux} we illustrate the LE Strehl ratio versus the field off-axis position. We observe a similar behavior than before. FEWHA with only $2$ iterations provides a better LE Strehl ratio than the LS reconstructor. With $4$ PCG iterations we can increase the quality for FEWHA further. If we increase the off-axis angle, the difference in quality of FEWHA compared to the LS reconstructor increases.

Summarizing, our quality analysis shows that FEWHA provides an excellent reconstruction quality when using only $2$ PCG iterations. When increasing the number of iterations to $4$ the quality slightly improves. In the upcoming section we study the computational performance of the algorithm on a Central Processing Unit (CPU).

\subsection{Computational performance}
Our previously conducted study reveals that the computational performance for FEWHA is better on a CPU than on a GPU\cite{stadler2020realtime}. This is mainly caused by the low number of FLOPs induced by the dual domain discretization and a matrix-free implementation. Moreover, the level of parallelism is low compared to a computationally intensive MVM approach. We run the parallel CPU implementation of FEWHA on Radon1\footnote{\url{https://www.oeaw.ac.at/ricam/hpc}}, the high performance cluster of the Radon Institute for Computational and Applied Mathematics in Linz. For our numerical simulations we use one compute node, which is equipped with two 8-core Intel Haswell processors (Xeon E5-2630v3, 2.4Ghz) and 128 GB of memory.

Without parallelization it would not be possible to meet the real-time requirements of a large AO system such as MAORY. FEWHA allows two types of parallelization, which we refer to as global and local parallelization. By global parallelization we understand the decomposition of the operators involved into layers $L$ or WFSs $W$ blocks. Local parallelization refers to parallelization inside these blocks. We implement FEWHA by combining OpenMP\footnote{\url{https://www.openmp.org/}} parallel regions for global parallelization with vector extensions for local parallelization. We apply explicit vectorization utilizing the Intel AVX2 \footnote{\url{https://software.intel.com/content/www/us/en/develop/documentation/cpp-compiler-developer-guide-and-reference/top/compiler-reference/intrinsics.html}} vector instructions for the discrete wavelet transform, the SH operator and the bilinear interpolation. The combination of the two strategies leads to a very efficient parallelization scheme\cite{stadler2020realtime}.

\begin{figure}
	\centering
	\includegraphics[width=1.0\textwidth]{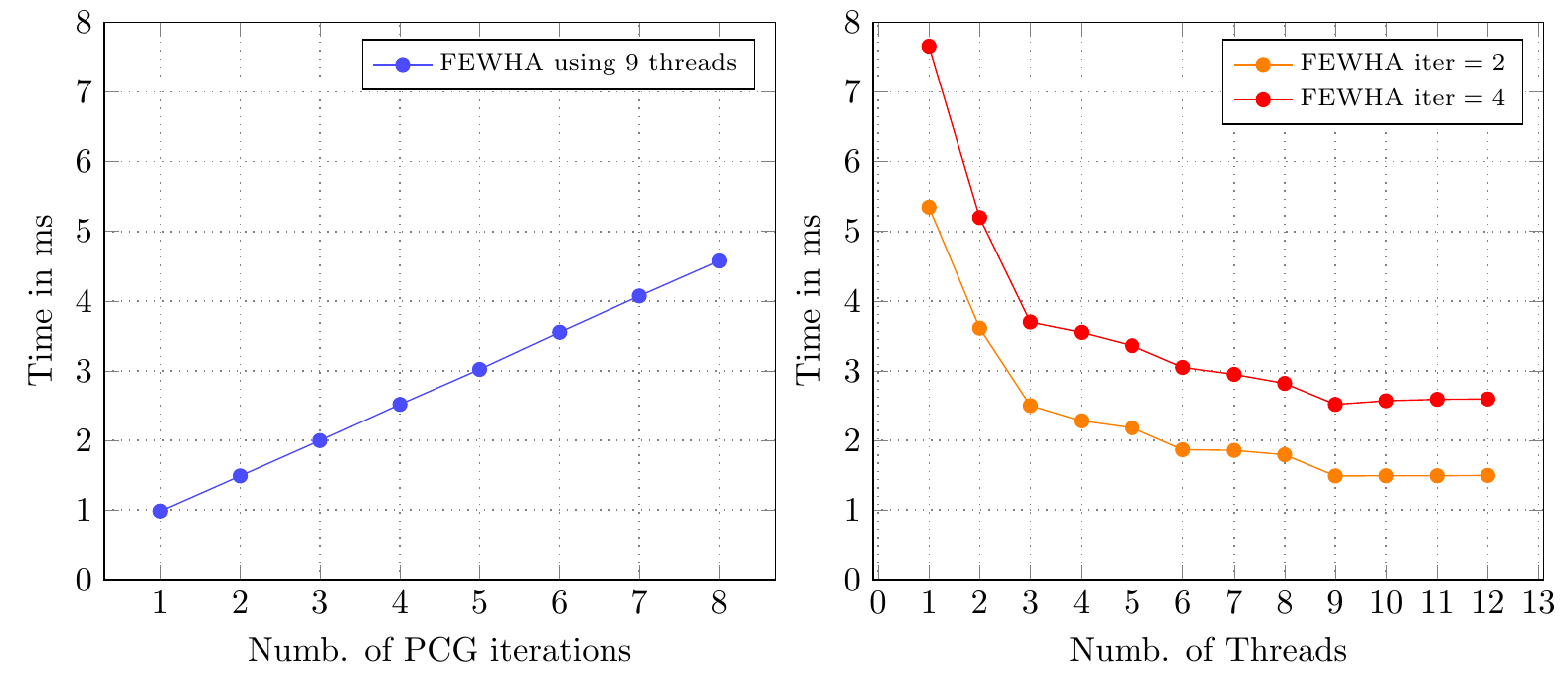}
	\caption{Scalability of FEWHA with number of PCG iterations (left) and number of threads (right) running on a multi-core CPU. The best performance, which balances quality and run-time, is obtained when using $2$ PCG iterations and $9$ threads for global parallelization.}
	\label{fig:scalability_run-time}
\end{figure}

In Figure~\ref{fig:scalability_run-time} we illustrate the hard real-time computational performance of FEWHA for MAORY. Note that our iterative algorithm does not require any soft real-time computational tasks. The method specific parameter values are given in Table~\ref{tab:methodParams}. The timings correspond to the average run-time for one out of $10000$ time steps. The left plot of Figure~\ref{fig:scalability_run-time} shows the parallel scalability of FEWHA with the number of PCG iterations. A linear relation between the number of iterations and the run-time is what we expect, because the PCG iterations are not parallelizable. In Section~\ref{sec:computational} we show that the wavelet reconstructor is able to provide a good reconstruction quality with only $2$ iterations. Utilizing $2$ iterations we observe from Figure~\ref{fig:scalability_run-time} that FEWHA performs the reconstruction within $1.48$~ms. The test runs are executed with $9$ parallel threads for global parallelization. In the right plot of Figure~\ref{fig:scalability_run-time} we illustrate the scalability with the number of threads. Here we use $2$ (orange) and $4$ (red) PCG iterations for the reconstruction. We observe that the best performance is obtained when using only $9$ threads for global parallelization. This number $9$ here corresponds to the number of WFSs.

\review{Note that we do not consider any pipelining for FEWHA in the timing measurements. Due to the latency from the data acquisition from the sensors, $s^{(i+1)}$ is not available at once. Hence, operations that do not require the complete vector can be started before the whole vector is available. This leads to an overlap of the time frame required for the data transfer and the one for calculation, and thus speeds up the overall computational performance. For FEWHA pipelining could be applied when computing the pseudo open loop measurements in Line~\ref{line:openLoop} of Algorithm~\ref{alg:augFEWHA}. The measurement vector $s^{(i+1)}$ is only required for the sum. Computing the sum of two vectors is perfectly pipelineable. Thus, as soon as elements of $s^{(i+1)}$ are accessible, the first elements of the result vector can be calculated.}

Besides being fast, the matrix-free approach of FEWHA leads to a significant reduction in memory requirements compared to the matrix-based MVM method. The units of memory required for the MVM for the MAORY setting are about $53$~GB\cite{StaAO4ELT}, mainly caused by storing the huge control matrix. For FEWHA the memory usage reduces to only $16$~MB. Moreover, this matrix-free implementation enables on the fly parameter updates.

\section{Conclusion}\label{sec:conclusion}
Direct solvers, such as the MVM method, have been used in the context of atmospheric tomography since the beginning. They are convenient to use, they are easy to implement and their application is well parallelizable and pipelineable. However, they have some non negligible drawbacks. First of all, the dimensions of \review{ESO's ELT} lead to a very large matrix. Storing one big matrix is memory consuming and it is very demanding to compute the generalized inverse in soft-real time and the matrix-vector multiplication in hard real-time. Fulfilling the real-time requirement is only possible with expensive hardware and a combination of parallelization and pipelining. Moreover, if certain parameters at the telescope or in the atmosphere change, the huge matrix has to be reassembled. Iterative methods do not require the demanding soft real-time tasks. They are fast and benefit from on the fly system updates. \review{In this paper, we continued our work regarding the iterative solver FEWHA\cite{stadler2020,RaSt2021} and studied the performance in terms of quality and run-time for the MAORY instrument via simulations in COMPASS.} The key features of the wavelet reconstructor are the matrix-free formulation and a reduced number of iterations induced by preconditioning and augmentation. These techniques considerably reduce the run-time and memory consumption, which is essential for ELT-sized problems. We showed via numerical simulations using COMPASS that the algorithm yields an excellent quality for a MAORY-like test configuration. Moreover, the parallel implementation on a multi-core CPU is able to fulfill the real-time requirements. We conclude that FEWHA is a very promising real-time reconstructor for the MAORY instrument of \review{ESO's ELT}.

\subsection* {Acknowledgments}
We want to thank Stefan Raffetseder for providing the interface between FEWHA and COMPASS and for his help to set up the MAORY simulation in COMPASS.
 
The project has received funding by the European Union’s Horizon 2020 research and innovation programme under the Marie Sk\l odowska-Curie Grant Agreement No.~765374, the Austrian Science Fund (FWF) F6805-N36 (Tomography in Astronomy) and the Austrian Research Promotion Agency (FFG) FO999888133 (Industrial methods for Adaptive Optics control systems). 

\bibliography{literature}
\bibliographystyle{unsrt}
\end{document}